%% file: J1544_timing_ver0.tex
\documentclass[twocolumn, twocolappendix]{aastex63}
\usepackage{amsmath}	
\usepackage{amssymb}	
\usepackage{scrextend}
\usepackage{lipsum,xcolor}
\usepackage{graphicx,midfloat,capt-of}

\shorttitle{First systematic study reporting the changes in eclipse cut-off frequency for PSR J1544+4937}
\shortauthors{Kumari et al.}

\begin{document}
\title{First systematic study reporting the changes in eclipse cut-off frequency for pulsar J1544+4937}
\correspondingauthor{Sangita Kumari}
\email{skumari@ncra.tifr.res.in}

\author[0000-0002-3764-9204]{Sangita Kumari}
\affiliation{National Centre for Radio Astrophysics, Tata Institute of Fundamental Research, S. P. Pune University Campus, Pune 411007, India}
\author[0000-0002-6287-6900]{Bhaswati Bhattacharyya}
\affiliation{National Centre for Radio Astrophysics, Tata Institute of Fundamental Research, S. P. Pune University Campus, Pune 411007, India}
\author[0009-0001-9428-6235]{Rahul Sharan}
\affiliation{National Centre for Radio Astrophysics, Tata Institute of Fundamental Research, S. P. Pune University Campus, Pune 411007, India}
\author[0000-0001-8801-9635]{Devojyoti Kansabanik}
\affiliation{National Centre for Radio Astrophysics, Tata Institute of Fundamental Research, S. P. Pune University Campus, Pune 411007, India}
\author[0000-0001-9242-7041]{Benjamin Stappers}
\affiliation{Jodrell Bank Centre for Astrophysics, Department of Physics and Astronomy, The University of Manchester, UK}
\author[0000-0002-2892-8025]{Jayanta Roy}
\affiliation{National Centre for Radio Astrophysics, Tata Institute of Fundamental Research, S. P. Pune University Campus, Pune 411007, India}

\begin{abstract}
We present results from a long-term monitoring of frequency dependent eclipses of the radio emission from PSR J1544$+$4937 which is a ``black widow spider'' millisecond pulsar (MSP) in a compact binary system. The majority of such systems often exhibit relatively long duration radio eclipses caused by ablated material from their companion stars. 
With the wide spectral bandwidth of upgraded Giant Metrewave Radio Telescope (uGMRT), we present first systematic study of temporal variation of eclipse cut-off frequency.
With decade-long monitoring of 39 eclipses for PSR J1544+4937, we notice significant changes in the observed cut-off frequency ranging from 343 $\pm$ 7 MHz to $\ge$ 740 MHz. We also monitored changes in eclipse cut-off frequency on timescales of tens of days and observed  a maximum change of $\ge$ 315 MHz between observations that were separated by 22 days. In addition, we observed a change of $\sim$ 47 MHz in eclipse cut-off frequency between adjacent orbits, i.e. on timescales of $\sim 2.9$ hours. We infer that such changes in the eclipse cut-off frequency depict an eclipse environment for the PSR J1544+4937 system {that} is dynamically evolving, where, along with the change in electron density, the magnetic field could also be varying. We also report a significant correlation between the eclipse cut-off frequency and the mass loss rate of the companion. This study provides the first direct evidence of mass loss rate affecting the frequency dependent eclipsing in a spider MSP.

\end{abstract}


\section{Introduction}
\label{sec:intro}
Black widow (BW) and Redback (RB) millisecond pulsars (MSPs), commonly classified as ``spider'' MSPs, are in compact binary systems with a low mass companion \citep[$\mathrm{M_{c}}< $ 0.05 M$_\odot$ for BWs, and 0.1 M$_\odot<\mathrm{M_{c}}< $ 0.9 M$_\odot$ for RBs,][]{roberts2012surrounded}. A majority of these systems exhibit relatively long duration \citep[e.g. $\sim 10\%$ for PSR J1959$+$2048,][]{Fruchter1988a} radio eclipses, where the ablated material from the companion star blocks the low-frequency radio waves from the pulsar \citep{polzin2020study}. The observed eclipses for the ``spider'' MSP systems are frequency dependent, where below a certain frequency (generally denoted as eclipse cut-off frequency, $\nu_{c}$) the pulsed signal disappears while the signal is detectable at higher frequencies during the low frequency eclipses. Although the exact nature of frequency dependence depends on the individual spider system, it has been observed that the eclipses are more pronounced at the lower frequencies compared to higher frequencies. 

Variable eclipses were seen for other black widow MSPs in the past. For example, for PSR J0024$-$7204J, Figure 8 and the related discussion in \cite{Freire200347tuc} indicates the possibility of time dependent eclipse cut-off frequency at 660 MHz (also discussed in \cite{Freire_2005_J0024}). Moreover for PSR J2051$-$0827, Figure 5 of \cite{PolzinJ2051} shows DM variation over time mapped for a decade, which may in turn indicate temporal changes in eclipse cut-off frequency. Although in \cite{PolzinJ2051} there is no explicit mention of such variability.

Because of the lack of wide bandwidth observing facilities available until recently (e.g. MeerKAT, Parkes UWB receiver, upgraded GMRT), the eclipse cut-off frequency is not precisely known (the reported eclipse cut-off frequency often have an ambiguity of $\sim$200 MHz) for the majority of the ``spider'' MSP systems \citep[an exception being PSR J1544+4937,][]{kansabanik2021unraveling}. 
This unavailability of precise eclipse cut-off frequency means that the temporal changes in the eclipse cut-off frequencies for any ``spider'' MSPs are not been monitored before the present attempt for PSR J1544+4937, reported in this paper. 

Discovered in the Fermi-directed search by the Giant Metrewave Radio Telescope (GMRT, \cite{swarup1991asp}), PSR J1544+4937 is a BW MSP with a spin period of 2.16 ms \citep{bhattacharyya2013gmrt}. This MSP is in a close binary system with an orbital period of 2.9 hours, and it is orbited by a low mass companion star with a minimum mass of 0.017 M$_{\odot}$ \citep{bhattacharyya2013gmrt}.

 Previous studies performed by \cite{bhattacharyya2013gmrt} with the legacy GMRT system having a bandwidth of 32 MHz  reported that eclipses were seen at 322 MHz for this MSP (where the signal from the pulsar was obscured for $\sim$13\% of the orbit) but no eclipse was observed at 607 MHz.  However, due to the lack of wide bandwidth observations, \cite{bhattacharyya2013gmrt} could not provide a more precise constraint on the cut-off frequency for PSR J1544+4937.

The first optical detection of the companion of MSP J1544+4937 is reported by \cite{tang2014identification}. 
A more recent optical study of this system by \cite{Mata_optical_J1544} revealed that a simple direct heating model can explain the observed light curves and found the inclination angle to be $\mathrm{47^{\circ +7}_{-4} }$. Along with this \cite{Mata_optical_J1544} also inferred that the companion is filling its Roche-lobe 
and estimated the Roche-lobe filling factor to be $>$ 0.96, which is consistent with the presence of the observed radio eclipses in this system.

 
 A detailed overview of the probable eclipse mechanisms is provided by \cite{Thompson1992}, which could explain the frequency dependent eclipsing in spider MSP systems. \cite{Thompson1992} emphasised that different eclipse mechanisms may be responsible for eclipses in different systems. For instance, cyclotron-synchrotron absorption is believed to be the major eclipse mechanism for  J1227$-$4853 \citep{Kudale2020}, PSR J1544$+$4937 \citep{kansabanik2021unraveling} and PSR J1810$+$1744 \citep{PolzinJ1810}, while scattering and cyclotron absorption are considered the primary mechanisms for PSR J2051$-$0827 \citep{PolzinJ2051}. Furthermore, stimulated Raman scattering has been suggested as the most plausible eclipse mechanism for PSR B1744$-$24A \citep{thompson1994physical}.
 However, as of now the eclipse properties have been investigated for only a handful of spider MSPs systems.

 Recent advancements in telescope bandwidths have made it possible to accurately determine and investigate the corresponding time dependent changes in the cut-off frequency. The upgraded GMRT \citep[uGMRT,][]{gupta2017upgraded,Reddy} is a perfect instrument to study the frequency-dependent eclipsing in spider MSP systems as it provides a wide frequency (120 MHz $-$ 1460 MHz) and wide bandwidth coverage (up to $\sim$ 400 MHz). \cite{kansabanik2021unraveling} reported the eclipse cut-off frequency for PSR J1544+4937 to 345$\pm$5 MHz, using observations with the uGMRT, which is an order of magnitude more precise than previous estimates.

In this paper, we present a decade long monitoring of frequency dependent eclipsing for PSR J1544+4937. The analysis of these changes enabled us to investigate the dynamical evolution of the eclipse medium for PSR J1544+4937.
The details of the observations and the data analysis are discussed in Section \ref{sec:obs}. In Section \ref{results} the results are presented. Section \ref{discussion} details the possible reason for the change in the cut-off frequency and Section \ref{conclusion} provides conclusion of this paper.

\input{Table1}

\section{Observations and data analysis}
\label{sec:obs}

This decade long monitoring of PSR J1544$+$4937 in the eclipse phase was performed using the GMRT \citep{swarup1991asp,gupta2017upgraded}, a radio interferometric array composed of 30, 45-meter dishes. 
The observations reported in this paper were performed in the phased array mode. 

The details of the observations used for this investigation are provided in Table {\ref{tab:Table1}}.
The initial observations, spanning 2011 to 2017, were conducted using the legacy GMRT 32 MHz bandwidth system \citep{roy2010real} at the central frequencies of 322 MHz and 607 MHz. Subsequent observations, from 2017 onwards, were performed using the uGMRT 200 MHz bandwidth system centered at 400 MHz and 650 MHz. The majority of the observations were conducted in the simultaneous dual-frequency mode by splitting the whole array into two sub-arrays with roughly equal number of antennas at band-3 (300$-$500 MHz) and band-4 (550$-$750 MHz). Coherent beam filterbank data at the best achievable time-frequency resolution were recorded (mentioned in Table \ref{tab:Table1}). In some of the uGMRT observing epochs (02 December 2019, 10 January 2020 and 15 February 2020), we used real time broadband radio frequency interference (RFI) removal system \citep{Buch_RFI_filter1,Buch_RFI_filter3,Buch_RFI_filter2} during our observations. Our observations spanned over a period of 10 years, covering a total of 39 eclipses. For two of the eclipses, observations were performed using the coherent de-dispersion technique \citep{coherent_dedispersion}.

To mitigate narrow-band and short-duration broad-band radio frequency interference (RFI), we employed the GMRT pulsar tool (gptool\footnote{\label{note1}\url{https://github.com/chowdhuryaditya/gptool}}) software. After the RFI mitigation, we corrected for the interstellar dispersion using the incoherent dedispersion technique and folded the resulting time series with the known radio ephemeris for PSR J1544+4937 from \cite{skumari2022}, utilizing the $\it{prepfold}$ task of $\it{PRESTO}$ \citep{ransom2002fourier}. The mean pulse profile was cross-correlated with a high signal-to-noise template profile from previous observations to obtain the observed times of arrival (TOAs) of the pulses. We note that there is no significant intra-band frequency evolution of the pulse profile for PSR J1544+4937. The TOAs were generated using the python script $\it get\_TOAs.py$ from $\it{PRESTO}$ \citep{ransom2002fourier}. We calculated the timing residuals, which is the difference between the observed and predicted TOAs, using the $\it{tempo2}$ software package \citep{hobbs2006tempo2}. The excess dispersion measure ($DM_{\mathrm{excess}}$) in the eclipse region (orbital phase $\sim 0.2 - 0.3$) introduces an extra time delay and can be determined using the relation \citep{LorimerKramer}:

\begin{equation}
\label{Excess_DM}
    DM_{\mathrm{excess}} = 2.4 \times 10^{-10} t_{\mathrm{excess}} (\mu s) f^{2} (MHz)
\end{equation}
where, $\mathrm{t_{excess}}$ is the excess time delay in the eclipse region in $\mu s$ and f is the observing frequency in MHz.
From $\mathrm{DM_{excess}}$ the electron column density in the eclipse medium ($N_{e}$) is computed using :
\begin{equation}
\label{N_e_determination}
\mathrm{N_{e}(cm^{-2}) = 3 \times 10^{18} \times DM (pc ~ cm^{-3})}   
\end{equation}

We used the following method to determine the frequency below which the pulsar signal disappears. Firstly, we divided the observing bandwidth into 15 MHz chunks and searched for pulsed signals within the eclipse region, which has been observed to lie between orbital phase 0.2 to 0.3 in previous studies done by \cite{bhattacharyya2013gmrt,kansabanik2021unraveling,skumari2022} for PSR  J1544+4937. 
A detection significance above 4$\sigma$ (refer Appendix \ref{Determining on and off Phase Bins}) in the eclipse region for a 15 MHz chunk in frequency was considered as detection. The uncertainty on the cut-off frequency value was estimated as half the width of the corresponding chunk in the frequency domain. Typically, the error in $\nu_{c}$ was $\sim$ 7 MHz for most of the eclipses. However, for a few eclipses where $\nu_{c}$ was between band-3 and band-4 (indicated by eclipsing in band-3 and detection in band-4), the error was estimated to be $\sim$ 35 MHz. 

\begin{figure*}[!htbp]
\begin{center}
\includegraphics[width=0.99\textwidth,angle=0]{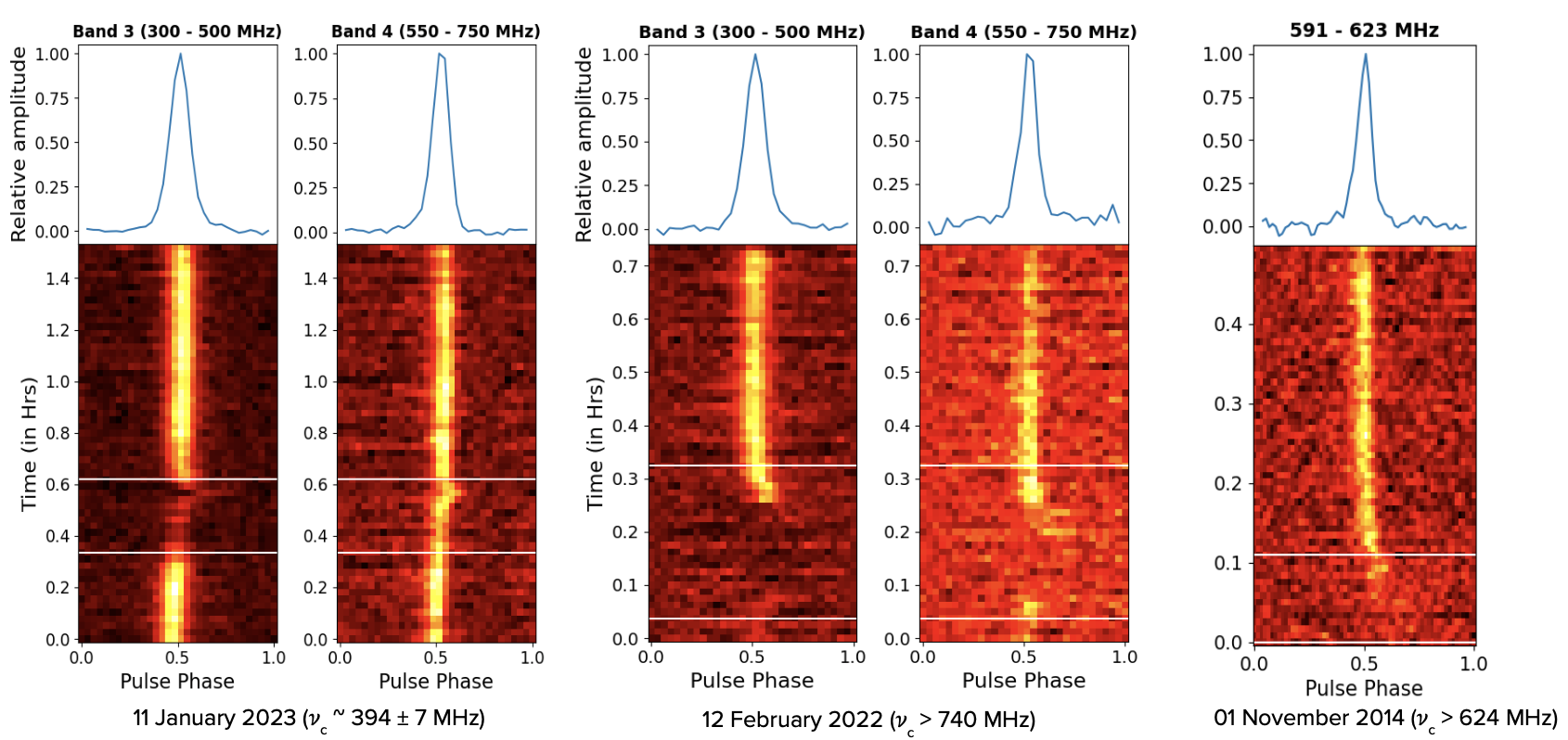}
\caption{Evolution of eclipse cut-off frequency with observing epochs is evident for the three epochs of observations shown in the different panels. The top plot of each panel shows the integrated pulse profile of the pulsar and the bottom plot of each panel shows time vs pulse phase plot. The orbital phase 0.2$-$0.3 is marked by the white lines to guide the eyes towards the eclipse region. The observing frequency range for each epoch is mentioned at the top of each panel and the corresponding cut-off frequency is given at the foot of each figure.}
\label{fig:cutoffchange}
\end{center}
\end{figure*}

In addition we calculated the flux density for the sample of eclipses during the non-eclipse phase (refer Appendix \ref{Determining on and off Phase Bins}). The non-eclipse phase is defined by excluding the orbital phase from 0.15 to 0.32, as the eclipse for this pulsar is confined to this orbital phase range. This definition of the non-eclipse phase excludes the region with increased delay at the eclipse boundary, as also observed in Figure \ref{fig:cutoffchange}.

We also estimated the mass loss rate of the companion using the relation, $\mathrm{\Dot{M_{c}} \sim \pi R_{E}^{2} m_{p} n_{e} V_{w}}$ \citep{Thompson1992,PolzinJ1810}, where $\mathrm{R_{E}}$ is the eclipse radius, $\mathrm{m_{p}}$ is the mass of the proton, $\mathrm{n_{e} = N_{e}/(2R_{E})}$, is the electron volume density, and $\mathrm{V_{w} = (U_{E}/(n_{e} m_{p}))^{1/2}}$, is the velocity of the material entrained in the pulsar wind. Here, $\mathrm{U_{E} = \Dot{E}/(4 \pi c a^{2})}$ represents the energy density of the pulsar wind, with $\Dot{E}$ being the spin-down energy and $\mathrm{a}$ being the distance between the pulsar and the companion. The mass loss rate has been calculated under the assumption that material is spherically symmetric around the companion and the orbital period does not change with time. The observed orbital period variation by \cite{skumari2022} for PSR J1544+4937 are of the order of $10^{-8}$ days which will have a negligible effect on the $\Dot{M_{c}}$ calculations. The eclipse radius is determined by estimating the number of sub-integrations in time that have signal to noise ratio (SNR) $\le$ 4 around the eclipse region (approximately $0.2 - 0.3$) for 300 MHz $-$ 345 MHz frequency chunk (refer Appendix \ref{Determining on and off Phase Bins}). This frequency chunk has been used, as the $\mathrm{\nu_{c}  \gtrsim  345 }$ MHz for all the eclipses observed for this MSP.

Table \ref{tab:Table3} lists the cut-off frequency for the sample of eclipses, the corresponding $\mathrm{N_{e}}$ and the $\mathrm{\Dot{M_{c}}}$ as well as the flux density in the non-eclipse phase calculated using the procedure mentioned in Appendix \ref{Determining on and off Phase Bins}.

\input{table3.tex}

\section{Results}
\label{results}
\subsection{A decade long mapping of eclipse cut-off frequency}
\label{A decade long mapping of eclipse cut-off frequency} 
We monitored the long term temporal behaviour of the eclipse cut-off frequency for PSR J1544+4937 which is presented in Table \ref{tab:Table3}.
Figure \ref{fig:cutoffchange} shows an example of different cut-off frequencies for 3 eclipses. We see that on 2023 January 11 and 2022 February 12, simultaneous dual frequency observations at band-3 and band-4 taken with the uGMRT 200 MHz bandwidth system revealed different values of the eclipse cut-off frequency ($\nu_{c}$), namely 394 $\pm$ 7 on 2023 January 11 and $\ge$ 740 MHz on 2022 February 12. Such drastic changes in $\nu_{c}$ is also seen with observations using the legacy GMRT 32 MHz bandwidth system. For example, Figure \ref{fig:cutoffchange} shows the time vs pulse phase plot for 2014 November 01, for which a full eclipse can nearly be seen in the whole 32 MHz bandwidth data centered at 607 MHz implying a cut-off frequency $\ge 624$ MHz. 

The long term temporal variation of cut-off frequency for PSR J1544+4937 is shown in upper panel of Figure \ref{fig:2}. The maximum observed value of $\nu_{c}$ is $\ge$ 740 MHz for a few eclipses (e.g. 2022 February 12, 2022 June 23, 2022 September 23), while the minimum value of $\nu_{c}$ is 343 $\pm$ 7 MHz on 2020 July 13. It can also be noted that the eclipse cut-off frequency is changing on timescales as short as a few days ($\sim 10$ days, between 2022 June 28 and 2022 July 08). We observed a maximum change of $\ge 315$ MHz between observations separated by 22 days (2023 March 04 and 2023 March 26).

We found a positive Spearman correlation coefficient\footnote{\label{note3}Estimated using SciPy python package} of 0.45 (with a corresponding probability value of 0.04) between the cut-off frequency and the corresponding Modified Julian Date (MJD), where we have considered only those eclipses for which the exact value of cut-off frequencies are known (marked in bold in Table \ref{tab:Table3}). This may mean that statistically the cut-off frequency is increasing with time (although the correlation is weak).

We investigated the change in the eclipse cut-off frequency between two consecutive orbits for three eclipses (on 2022 December 24, 2022 December 30, and 2023 January 28). We did not find any change in $\nu_{c}$ on 2022 December 24, but we did notice a change in the value of $\nu_{c}$ between the two consecutive orbits on 2022 December 30 (see in Table \ref{tab:Table3}). The observed change in $\nu_{c}$ between consecutive orbits for 2022 December 30 implies that the eclipse environment is changing on an hour's timescale. However, for the eclipses on 2023 January 28, we were unable to estimate the eclipse cut-off frequency for the second eclipse covered. Within the eclipse region, the signal was present in the latter portion of band-3, but no detection was observed in band-4. This anomaly may be attributed to the change in the spectral index within two consecutive eclipses, where the flux density at band-4 for the second eclipse is lower than that for the first eclipse, while the flux densities at band-3 are comparable in both eclipses. The flux density values are given in Table \ref{tab:Table3}.

Moreover, our findings indicate marginal differences in the eclipse cut-off frequencies compared to those reported by \cite{kansabanik2021unraveling}. Specifically, \cite{kansabanik2021unraveling} noted a consistent cut-off frequency of 345 MHz for PSR J1544+4937 across three observation eclipses (2018 February 6, 2018 April 17, 2018 May 7), while our analysis reports slightly varied cut-off frequencies for these eclipses. These differences in the cut-off frequency values are attributed to differences in the method and threshold employed for the determination of the cut-off frequency along with RFI mitigation techniques.


\subsection{A decade long mapping of electron column density in the eclipse region}
\begin{figure*}[!htbp]
\begin{center}
\includegraphics[width=0.99\textwidth,angle=0]{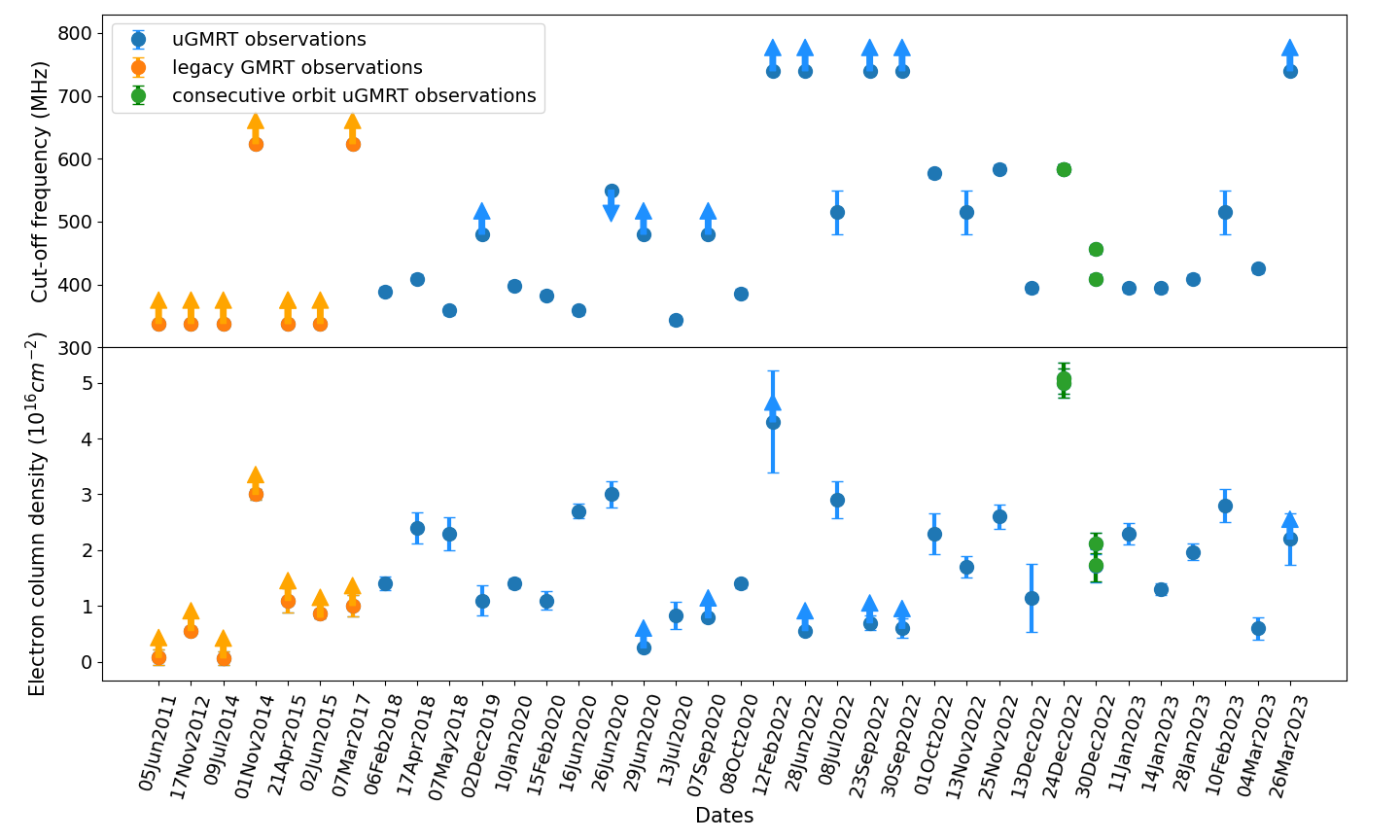}
\caption{Upper panel: The long term temporal variation of eclipse cut-off frequency for PSR J1544+4937. The upward arrow depicts that the cut-off frequency on the respective eclipses is only the lower limit as in these complete eclipse is observed. Whereas, the downward arrow depicts that the cut-off frequency for that eclipse is the upper limit. For 2022 July 08, 2022 November 13 and 2023 February 10 the cut-off frequency is between band-3 and band-4 and hence the error of $\pm$ 35 MHz.\\
Lower panel: The long term variation of  electron column density in the eclipse region. The upward arrow depicts that the column density on the respective eclipses is only the lower limit as it is estimated at the eclipse boundary. The errors on the $N_{e}$ values are calculated from respective errors on TOAs of the pulses measured at the eclipse boundary or superior conjunction.}
\label{fig:2}
\end{center}
\end{figure*}

We also studied the long term variation of electron column density ($N_e$) in the eclipse region for PSR J1544+4937 (lower panel of Figure \ref{fig:2}). The estimated values of $N_{e}$ are given in Table \ref{tab:Table3}. The maximum value of $N_{e}$ estimated is 5$\times 10^{16} \mathrm{cm^{-2}}$ on 2022 December 24, where as the minimum value of $N_{e}$ is 6.6$\times 10^{15} \mathrm{cm^{-2}}$ on 2023 March 04. For certain eclipses $N_{e}$ could only be estimated at the eclipse boundaries providing us with a lower limit, which is indicated by the upward arrows in the lower panel of Figure \ref{fig:2}. 
For these eclipses complete disappearance of pulsed signal in the eclipse phase for both band-3 and band-4 makes it impossible to estimate the maximum value of $N_{e}$ near the superior conjunction.
We also note that there is no systematic time dependent trend observed for variation of $N_{e}$.

We also investigated whether $N_{e}$ changes between two consecutive eclipses on 2022 December 24, 2022 December 30 and 2023 January 28. From Figure \ref{fig:2}, it can be noted that on 2022 December 24, the electron column density in the eclipse region for two consecutive orbits is almost the same which aligns with the absence of any change in the cut-off frequency. On the 30th December 2022, there is a slight difference in the electron column density for two consecutive orbits, although considering the error, this is not significant. For these consecutive eclipses we also noted a change in the eclipse cut-off frequency (refer Section \ref{A decade long mapping of eclipse cut-off frequency}). On, 2023 January 28 there is no detectable change of $N_{e}$ between two consecutive eclipses (refer Table \ref{tab:Table3}).

From Table \ref{tab:Table3}, it is evident that for the eclipses on 2014 July 09 and 2014 June 05 the error on $N_{e}$ is greater than the value of $N_{e}$ itself. The error on $N_{e}$ for these two eclipses are similar to the errors for other eclipses but the observed value of $N_{e}$ is smaller at the eclipse boundary compared to other eclipses.

\subsection{Mass loss rate of the companion}
The mass loss rate $\mathrm{\Dot{M_{c}}}$ for the eclipses in our sample ranges between $\mathrm{25.3 \times 10^{-14} }$ M$_\odot$/yr
$\mathrm{ \ge \Dot{M_{c}} \ge 2.24 \times 10^{-14}}$ M$_\odot$/yr. From Table \ref{tab:Table3}, it can be seen that for a few eclipses we do not have the measurement of the eclipse radius and hence the $\mathrm{\Dot{M_{c}}}$, as for these eclipses the full eclipse phase was not covered. For the $\mathrm{\Dot{M_{c}}}$ calculations, we used $\mathrm{M_{c}}$ to be 0.02 \(M_\odot\), which was derived using the mass function value from timing \citep{skumari2022} for the inclination angle value of 47$^{\circ}$ \citep{Mata_optical_J1544}. We assumed that the orbit of the companion is circular for conversion of eclipse duration in minutes to solar radius. We used the average value of $N_{e}$ in the eclipse region for $\mathrm{\Dot{M_{c}}}$ calculations.

\section{Probing the changes in the eclipse cut-off frequency}
\label{discussion}

This paper reports the first systematic study depicting the temporal changes in the eclipse cut-off frequency for any spider MSP, where we noted that the eclipse cut-off frequency for different eclipses of observations varies between $\nu_{c} = 343 \pm 7 $ to $\nu_{c}\ge 740$ for PSR J1544+4937. Such changes in the frequency dependent eclipsing imply temporal evolution of the eclipse environment. To probe this further we have considered the possible eclipse mechanisms prescribed by \cite{thompson1994physical} to explain the observations using values of $\nu_{c}$ and N$_{e}$ as listed in Table \ref{tab:Table3}. An overview of the possible eclipse mechanisms is presented in Appendix \ref{detailed overview of the eclipse mechanism}. 

Considering synchrotron absorption by the trans relativistic non-thermal free electrons as an eclipse mechanism, we found that frequency dependent eclipsing can be explained \citep[similar to what was found by,][]{bhattacharyya2013gmrt,kansabanik2021unraveling}. Therefore, to probe the changes in the eclipse environment, we investigated the observed temporal changes in the eclipse cut-off frequency (reported in Section \ref{A decade long mapping of eclipse cut-off frequency}) using synchrotron absorption as the major eclipse mechanism. The optical depth for synchrotron absorption is given by \citep{thompson1994physical}:
\begin{equation}
\label{synchrotron}
    \tau_{abs} = \left( \frac{3^{\frac{(p+1)}{2}} \Gamma (\frac{3p+2}{12}) \Gamma(\frac{3p+22}{12})}{4}\right) \left( \frac{\sin \theta}{m} \right)^{\frac{p+2}{2}} \frac{n_{o} e^{2}}{m_{e}c\nu} L
\end{equation}
where $\theta$ is the angle of the magnetic field lines with our line of sight, L is the absorption length, $p$ is the non-thermal electron power law index index ($n(E) = n_{o}E^{-p}$), $n_{o}$ is the non-thermal electron density which is assumed to be 1$\%$ of the total density, $m_{e}$ is the mass of the electron, m is the cyclotron harmonic at frequency $\nu$ ($\nu/\nu_{o}$, where $\nu_{o}= eB/2\pi m_{e}c$), $e$ is the charge on the electron and $c$ is the speed of light. Figure \ref{cutoff_parameter} presents the change in the cut-off frequencies with the change of different parameters according to Equation \ref{synchrotron}, which is discussed in detail in Appendix \ref{detailed overview of the eclipse mechanism}. We demonstrated that different cut-off frequencies can be attained by a change of a single parameter, keeping others as constant.
It is apparent that a cut-off frequency more than 740 MHz as observed by us, is also reproducible for certain combinations of $p$, $\theta$, $B$ and $N_{e}$.

\subsection{Electron column density contribution to eclipse cut-off frequency changes}
\label{Electron column density contribution to eclipse cut-off frequency changes}

From Figure \ref{fig:2} it is evident that for some of the eclipses we only have a lower limit of $N_{e}$. 
For these eclipses we predicted the maximum value of $N_{e}$ possible using Equation \ref{synchrotron} for a given cut-off frequency with the assumption that all other parameters remain constant. 

We consider a simple scenario where the changes in the cut-off frequency are assumed to be solely produced by the changes in the electron column density in the eclipse medium. This assumption implies that the magnetic field, the angle between the magnetic field lines with our line of sight ($\theta$), and electron energy spectral index (p) remain constant and do not vary between the eclipses. 
We considered the magnetic field to be equal to the characteristic magnetic field ($B_{E}\sim$ 10 $\mathrm{Gauss}$), calculated using the pressure balance between pulsar wind energy density ($U_{E}=\frac{\dot{E}}{4\pi ca^{2}}$) and the stellar wind energy density of the companion ($\frac{B_{E}^{2}}{8\pi}$), where a is the distance between the pulsar and the companion ($a \sim 1.37 R _{\odot}$) and c is the speed of light.

We performed a curve fitting analysis on data from eclipses where both the exact cut-off frequency ($\nu_{c}$) and the electron column density ($N_{e}$) were available, to determine the best fit curve (for a $p$ and $\theta$) that aligns with the theoretical prediction from Equation \ref{synchrotron} (see Appendix \ref{Calculation of best fit values of p and theta}). The magnetic field strength was held constant at 10 $\mathrm{Gauss}$ and the optical depth ($\tau$) value is taken to be 1 throughout the analysis. 
The best fit \footnote{\label{note2}To accomplish the parameter optimization, we employed the $curve\_fit$ function from the scipy package in Python} gives  $p \sim 2\pm$ 1.19 and $\theta \sim 0.17 \pm 0.19$ $\mathrm{radians}$ \citep[restricting the allowed value of $p$ between 2$-$7][]{Dulk&Marsh}. 

Table \ref{tab:Table3} highlights the eclipses for which exact values of $N_{e}$ and the cut-off frequency were available. 
Using these optimal values of $\theta$, $p$, $B$ and taking optical depth $\tau$ equals to 1, the predicted values of $N_{e}$, that are required at the superior conjunction calculated using Equation \ref{synchrotron} are given in Table \ref{table2}. It is evident that in order to account for the observed change in cut-off frequency solely due to the change in electron density in the eclipse medium, a very sharp increase in the electron column density at the superior conjunction is required for 2020 June 29, 2022 June 28, and 2022 September 23.

We also found a  moderate correlation (Spearman's correlation coefficient\footnote{\label{note3}Estimated using SciPy python package} $\sim 0.55$ with probability value of 0.01) between the variation in electron column density and cut-off frequency in our data set. 
This correlation is found using the eclipses where the exact value of the cut-off frequency and electron column density is known (marked in bold in Table \ref{tab:Table3}). We also estimated the correlation using all the eclipses and got a value of 0.39 with probability value of 0.01.
The observed correlation suggests that variations in the eclipse cut-off frequency may not be exclusively due to fluctuations in electron column density, and other factors, such as magnetic field strength in the eclipse medium, could also contribute to these variations.

\begin{table*}[!htbp]
\begin{center}
\caption{Table showing the required electron column density at the superior conjunction (orbital phase $\sim$ 0.25), if the cut-off frequency changes were only produced by changes in the electron column density ($N_{e})$.}
\label{tab:Table2}
\vspace{0.3cm}
\label{table2}
\begin{tabular}{|l|l|l|l|l|l|l|l|}
\hline
Eclipse  $^{\ast}$     & Cut-off frequency $^{\dagger}$ &Orbital phase $^{\ast\ast}$ & Predicted $N_{e}$ $^{a}$ & Observed $N_{e}$ $^{b}$ \\
\hline
17 November 2012 & $\ge$ 338 & 0.19 & 7.9$\times10^{15}$ & (5.6$ \pm 0.5) \times10^{15}$ \\
\hline
09 July 2014 & $\ge$ 338 & 0.18 & 7.9$\times10^{15}$ & (7.05$\pm 11.9) \times10^{14}$\\
\hline
01 November 2014 & $\ge$ 624 & 0.28 & 4.9$\times10^{16}$ & (2.9$\pm 0.2) \times10^{16}$  \\
\hline
21 April 2015 & $\ge$ 338 & 0.18 & 7.9$\times10^{15}$ & (1.1$\pm 0.08) \times10^{16}$\\
\hline
02 June 2015 & $\ge$ 338 & 0.30 & 7.9$\times10^{15}$ & (8.06$ \pm 0.7) \times10^{15}$\\
\hline
07 March 2017 & $\ge$ 624 & 0.26 & 4.9$\times10^{16}$ & (1.01$\pm 0.1) \times10^{16}$ \\
\hline
29 June 2020 & $\ge$ 480 & 0.28 & 2.2$\times10^{16}$ & (2.53$\pm 0.6) \times10^{15}$\\
\hline
07 September 2020 & $\ge$ 480 & 0.30 & 2.2$\times10^{16}$ & (7.9$\pm 0.6) \times10^{15}$ \\
\hline
12 February 2022  & $\ge$ 740 & 0.28 & 8.3$\times10^{16}$ & (4.3$\pm 0.05) \times10^{16}$ \\
\hline
28 June 2022  & $\ge$ 740 & 0.32 & 8.3$\times10^{16}$ & (1.06$\pm 0.06) \times10^{16}$  \\
\hline
23 September 2022  &$\ge$ 740 & 0.30 & 8.3$\times10^{16}$ & (7.02$\pm 1.2) \times10^{15}$ \\
\hline
30 September 2022  &$\ge$ 740& 0.30 & 8.3$\times10^{16}$ & (6.10$\pm 1.7) \times10^{16}$ \\
\hline
26 March 2022 &$\ge$ 740& 0.28 & 8.3$\times10^{16}$ & (2.28$\pm 0.4) \times10^{16}$ \\
\hline
\end{tabular}
\end{center}
$^{\ast}$ : Epoch of the eclipse observation\\
$^{\dagger}$ : The observed eclipse cut-off frequency \\
$^{\ast\ast}$ : {\bf The orbital phase up-to which the pulsar signal is detected in the eclipse region. Thus the lower limit on the value of $N_{e}$ is obtained at this the orbital phase}\\
$^{a}$: Predicted $N_{e}$ at the superior conjunction (orbital phase $\sim$ 0.25), taking $B \sim 10 \mathrm{Gauss}$, $\theta \sim 0.17 \mathrm{radians}$ and $p \sim$ 2 (see Equation \ref{synchrotron}) and considering the lower limit of cut-off frequency\\
$^{b}$: The observed $N_{e}$ calculated at the orbital phase given in column 3\\

\vspace{1cm}
\end{table*}

\subsection{Magnetic field contribution to the eclipse cut-off frequency changes}

A decade-long timing study of PSR J1544+4937 by \cite{skumari2022} revealed secular variation of its orbital period.
The change in orbital period is studied along with the variations of the epoch of ascending node, as done for other BW pulsars \citep{ng2014high,shaifullah201621}. The observed changes may be attributed to variations in the gravitational quadrupole moment of the companion \citep{skumari2022} which could be produced by strong magnetic fields generated by stellar convection caused by tidal forces and the rapid rotation of the companion in the compact orbits of the BW systems. The alteration of the quadrupole moment changes the magnitude of the gravitational force between the pulsar and the companion, which results in small quasi-periodic oscillations in the orbital period \citep{Applegate1992}. This mechanism is also believed to be responsible for orbital period changes or epoch of ascending node variations in other BW MSPs systems \citep[e.g.][]{lazaridis2011evidence,applegate1994orbital}.

We found a moderate correlation (Spearman's correlation coefficient\footnote{\label{note3}Estimated using SciPy python package} $\sim$ 0.55 with probability value of 0.03) between the variation of eclipse cut-off frequency and the epoch of ascending node in our data set using those eclipses where the exact value of the cut-off frequency is known (marked in the bold in Table \ref{tab:Table2}). This moderate correlation may indirectly indicate that the magnetic field of the companion which is responsible for the variation of ascending node could also be responsible for the cut-off frequency variation.  

\subsection{Flux density contribution to eclipse the cut-off frequency changes}
The variations in the observed flux density may be caused by changes in the interstellar medium or could be intrinsic to the pulsar. Using eclipses observed with uGMRT system we found some negative weak correlation (Spearman's correlation coefficient\footnote{\label{note3}Estimated using SciPy python package} $\sim$ $-$ 0.36 with probability value of 0.03) between the variation in flux density of the pulsar at band-3 in the non-eclipse phase and the cut-off frequency.

Assuming that the fluctuations in the observed flux density are intrinsic to the pulsar (not originating from the interstellar medium), we infer that the outgoing flux from the eclipse medium would be proportional to the incident flux (flux entering the eclipse medium), provided the values of $N_{e}$, $p$, $\theta$, and $B$ remain constant. Consequently, for a constant sensitivity of the telescope at a given frequency, the intrinsic changes in the flux density of the pulsar may result in variation in the eclipse cut-off frequency. Negative correlation between the variation of flux density and the eclipse cut-off is expected from this argument.

We also examined the correlation between $N_{e}$ and the flux density of the pulsar at band-3 but found no evidence of the same in our data set.

\subsection{Mass loss rate contribution to the eclipse cut-off frequency changes}
Mass loss in the companion star can be attributed to three primary reasons. Firstly, the irradiation of the companion star by the pulsar wind or $\gamma-$rays from the intra-shock binary region \citep[observed for few spider MSPs,][]{Intrabinary_shock_J1311-3430,Intrabinary_shockJ2241-5236}, can heat the star causing it to expand resulting in evaporative mass loss \citep{companion_winds_nature,companion_wind_nature_2}. Secondly, due to heating effects, the companion may become bloated and potentially reach a point where it fills its Roche lobe. Overflowing from the Roche lobe can result in mass loss. Thirdly, the pulsar wind can directly ablate material from the companion.

In the case of PSR J1544+4937, optical studies \citep{Mata_optical_J1544} have confirmed that the pulsar is filling its Roche lobe, indicating that Roche lobe overflow could be a possible source of mass loss in this system. The $\mathrm{\Dot{E}/a^{2}}$ values for spider MSPs are notably higher compared to those of other MSPs\footnote{\label{note1}\url{https://www.atnf.csiro.au/research/pulsar/psrcat/}, assuming an orbital inclination angle of $60^{\circ}$}, which suggests strong ablation of the companion in spider MSPs and, consequently, mass loss. For PSR J1544+4937, the $\mathrm{\Dot{E}/a^{2}}$ value is 0.06 $\mathrm{\times 10^{35}}$ erg/s$\mathrm{R_\odot^{2}}$ \citep[calculated using an inclination angle of $47^{\circ}$,][]{Mata_optical_J1544}, which is similar to other spider MSPs that exhibit eclipses \citep[Table 1 of][]{Kudale2020}.

We observed a strong correlation (Spearman's correlation coefficient\footnote{\label{note3}Estimated using SciPy python package} $\sim$ 0.80 with probability value of $1.81 \times 10^{-6}$) between $\mathrm{\Dot{M_{c}}}$ and the cut-off frequency, using all the available eclipse data. This suggests that a higher mass loss rate in the companion corresponds to a higher cut-off frequency. Similarly, using only the epochs where precise values of the eclipse cut-off frequency (marked in bold in Table \ref{tab:Table3}), electron density, and $\mathrm{\Dot{M_{c}}}$ were known, we also found a strong correlation (Spearman's correlation coefficient\footnote{\label{note3}Estimated using SciPy python package} $\sim $0.72 with probability value of 0.001).

Previous studies report $\mathrm{\dot{M_{c}}}$ at a similar order of magnitude as observed for PSR J1544+4937. For instance, for PSR J2051$-$0827, $\mathrm{\dot{M_{c}} = 10^{-14}}$ M$_\odot$/yr \citep{stappers1996probing}, for PSR J1810+1744, $\mathrm{\dot{M_{c}} = 9 \times 10^{-12}}$  M$_\odot$/yr \citep{PolzinJ1810} and for PSR J1959+2048 and PSR J1816$+$4510, $\mathrm{\dot{M_{c}}}$ is $10^{-12}$ and $2 \times 10^{-13}$ M$_\odot$/yr, respectively  \citep{polzin2020study}. However, no previous study has examined the temporal evolution of $\mathrm{\dot{M_{c}}}$. A significant correlation between $\mathrm{\dot{M_{c}}}$ and the cut-off frequency suggests that $\mathrm{\dot{M}_{c}}$ could be the primary factor contributing to the observed frequency-dependent eclipsing.  

\section{Conclusion}
\label{conclusion}
This study presents the first systematic examination, illustrating the time-dependent alterations in the eclipse cut-off frequency for a spider MSP. After considering different eclipse mechanism for the observed eclipses for PSR J1544+4937 we concluded that they are caused by synchrotron absorption from the trans-relativistic free electrons present in the eclipse medium. 
Analysis of the long term variation of the eclipse cut-off frequency in PSR J1544+4937 led to the inference that the observed changes cannot be attributed to a single parameter but rather result from a complex interplay of multiple contributing factors.
Our investigation also revealed a moderate correlation between the variation in electron column density and cut-off frequency as well as between the variation in the epoch of the ascending node and cut-off frequency. Moreover, we noticed a negative weak correlation between the observed flux density and the eclipse cut-off frequency. We also observed a very significant correlation between the mass loss rate of the companion and the eclipse cut-off frequency. Presenting the first direct evidence that the mass loss rate from the companion could be affecting
the frequency dependent eclipsing. Finally, we suggest that changes in the magnetic field, the observed electron column density, fluctuations in the pulsar's intrinsic flux density and mass loss rate of the companion could be the contributing factors to the observed variations in the eclipse cut-off frequency of PSR J1544+4937. 
Although this study presents an investigation using the largest sample of the eclipses published so far for any spider MSP, yet an even larger sample would provide more stringent probe to the observed variations in the eclipse cut-off frequency.


\begin{acknowledgments}
We acknowledge support of the Department of Atomic Energy, Government of India, under project no.12-R\&D-TFR-5.02-0700. The GMRT is run by the National Centre for Radio Astrophysics of the Tata Institute of Fundamental Research, India. 
We sincerely thank the anonymous reviewer whose insightful comments and suggestions have significantly improved the paper.
\end{acknowledgments}

\bibliography{citation.bib}{}
\bibliographystyle{aasjournal}

\appendix

\section{Determining ON/OFF Phase Bins for measurement of eclipse cut-off frequency, eclipse radius and non-eclipse phase flux density}
\label{Determining on and off Phase Bins}
In the baseline subtracted data cube (time, frequency, phase-bin), OFF and ON phase bins both follow Gaussian random distribution with the same standard deviation but with different mean. Hence, for determining ON and OFF phase bins, a parameter, H = (sum of N number of samples)/(standard deviation $\times \sqrt{tf}$), is calculated for each phase bin, where N (= tf) is the number of samples for a given timestamp (t) and chunk in frequency (f). More detailed overview of the above procedure can be found in Sharan et al. (under preparation). In the following we describe the method used for the calculation of eclipse cut-off frequency and non-eclipse phase flux density.
\begin{itemize}

\item Eclipse cut-off frequency determination: 
For the cut-off frequency calculations, in the expression for H, we used t corresponding to number of sub-integrations in time in orbital phase, $\sim$ 0.2$-$0.3, covering the eclipse and f corresponding to number of sub-bands in frequency for 15 MHz chunk.
The phase bins which have the value of H greater than 4, were labeled as ON, otherwise OFF.

\item Eclipse radius measurement: 
For the eclipse radius estimation, in the expression of H, we have used t corresponding to a sub-integration in time ($\sim$ 1.5 mins) and f corresponding to the number of sub-bands in frequency for 300 MHz $-$ 345 MHz chunk (since $\mathrm{\nu_{c}  \gtrsim  345 }$ MHz for all the eclipses observed).
The phase bins which have the value of H greater than 4 were labeled as ON, otherwise OFF. The SNR of ON bins is calculated for each sub-integration in time following \cite{LorimerKramer}. The eclipse radius is determined by estimating the number of sub-integrations whose SNR is $\le$ 4 near the eclipse region, in the SNR versus time plot. 

\item
Flux density calculation: For the flux density estimations, in the expression for H, we have used, t corresponding to the number of sub-integrations in time for the non-eclipse phase ($\sim$ 0.15$-$0.32) and f corresponding to the number of sub-bands in whole 200 MHz observing band. The phase bins which have the value of H greater than 5, were labeled as ON, otherwise OFF. Subsequently, the SNR of all the ON bins is calculated \citep{LorimerKramer}. We then used the radiometer equation for the phased array mode to find the flux density \citep{LorimerKramer}, given by $\mathrm{S_{mean} = \frac{SNR~T_{sys}}{G n_{a}\sqrt{n_{p}\Delta fT}}\sqrt{\frac{W}{P-W}}}$, where $\mathrm{T_{sys}}$ represents the system temperature, G is the gain, $\mathrm{n_{a}}$ represents the number of antennas, $\mathrm{\Delta f}$ is the bandwidth, T is the observing time, W is the pulse width, and P is the spin period of the pulsar. We note that scatter broadening, which is 0.0019 ms, is negligible compared to pulse width. The flux density in the non-eclipse phase is calculated for the regions that are unaffected by sudden disappearance of pulse signal known as short eclipse \citep{bhattacharyya2013gmrt}.
The flux densities reported in Table \ref{tab:Table3} are estimated at the central frequency of 400 MHz with a 200 MHz bandwidth for the majority of the epochs. However, these flux density estimates may get affected if there is a variation in the in-band spectra with time.
\end{itemize}

\section{Possible eclipse mechanism for PSR J1544+4937}
\label{detailed overview of the eclipse mechanism}
The radio signal from the pulsar will not be detected if the frequency of the radio waves from the pulsar is less than the plasma frequency of the eclipsing medium. We calculated the electron volume density of   $\mathrm{n_{e} \sim 10^{5} ~cm^{-3}}$ ($N_{e}/L$, where L is the absorption length taken to be 1 \(R_\odot\)) in the eclipsing medium, considering all the observing eclipses. Using this value of $\mathrm{n_{e}}$ the plasma frequency is $\mathrm{f_{p}  = 8.5~ (\frac{n_{e}}{cm^{-3}})^{1/2} ~ kHz } \ge 2.6 ~\mathrm{MHz}$. The measured plasma frequency is significantly lower than the eclipse cut-off frequency ($\nu_{c}$). Therefore, plasma frequency cut-off could not be the viable eclipse mechanism. For certain eclipses, where only a lower limit of electron column density is available, the required changes in electron volume density at superior conjunction to account for the observed $\nu_{c}$ are so extreme that they are not realistically achievable.
Eclipse due to refraction can be ruled out as the required time delay of the pulses is 10$-$100 $ms$ \citep{thompson1994physical} for it to explain the observed eclipse, whereas the observed time delays are in $\mu s$.

The possibility of an eclipse caused by free electron scattering in the eclipsing medium can also be eliminated. To explain the $\nu_{c}$ of 345 MHz, \cite{kansabanik2021unraveling} discarded pulse broadening due to scattering as the major eclipse mechanism. Since scattering is frequency-dependent ($\sim \nu^{-4}$), therefore to explain the higher $\nu_{c}$ we also reject scattering of radio waves as the major eclipse mechanism.

Free-free absorption as a possible eclipse mechanism can also be ruled out as either very low temperatures ($\mathrm{T < 1000 ~K}$) or very high clumping factors are required ($\mathrm{f_{cl} = 10^{9}}$) in the eclipsing medium. The pulsar radiation is itself intense enough to heat the plasma beyond this required temperature value and also such a high value of a clumping factor is not feasible in the eclipse environment \citep{clumpingfactor}. 

Induced Compton scattering cannot be considered as the primary mechanism for the eclipse, as the calculated optical depth is less than one. 

\begin{figure*}[!htbp]
 \includegraphics[width=.52\linewidth]{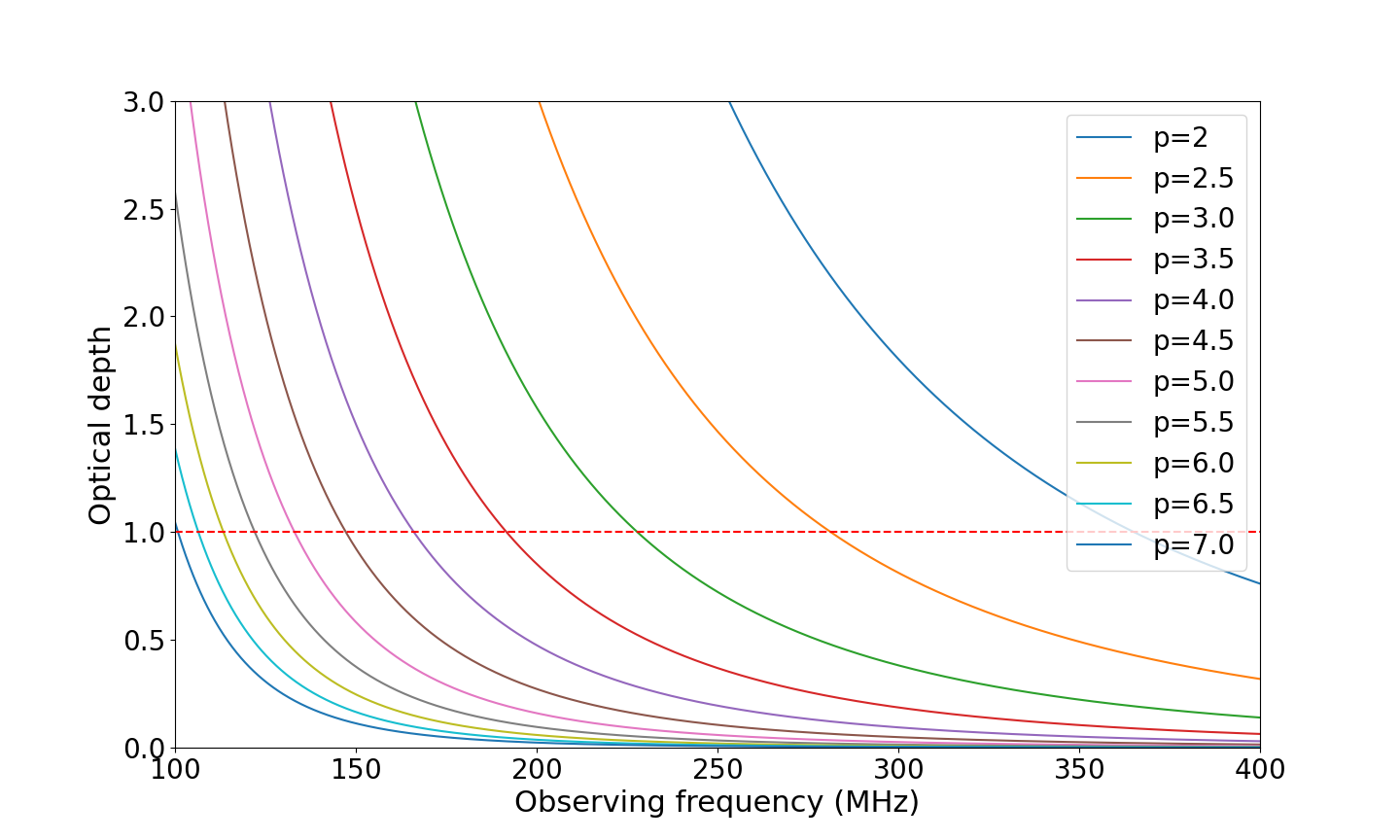} \hfill
 \includegraphics[width=.52\linewidth]{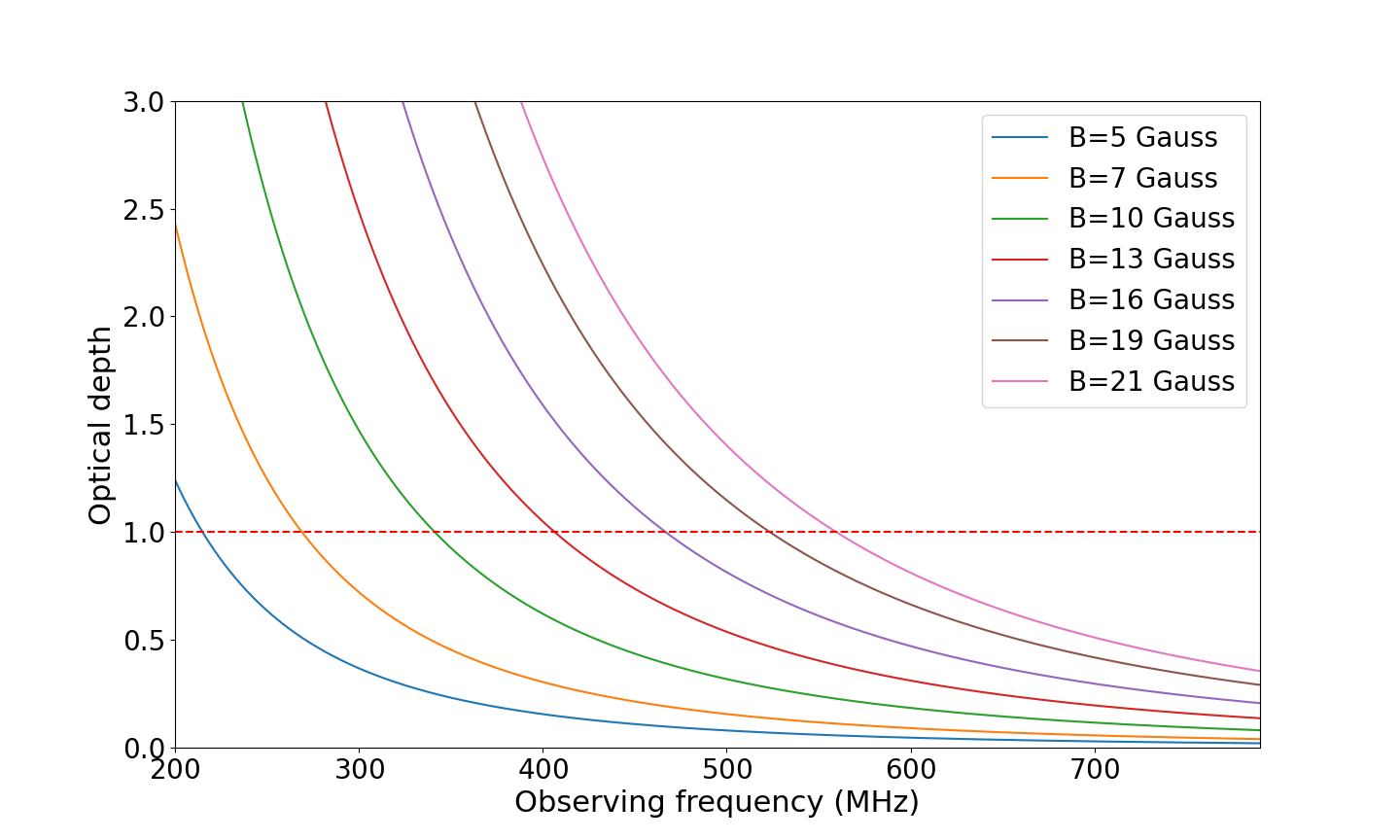} \\
 \vspace{1mm}
 \includegraphics[width=.52\linewidth]{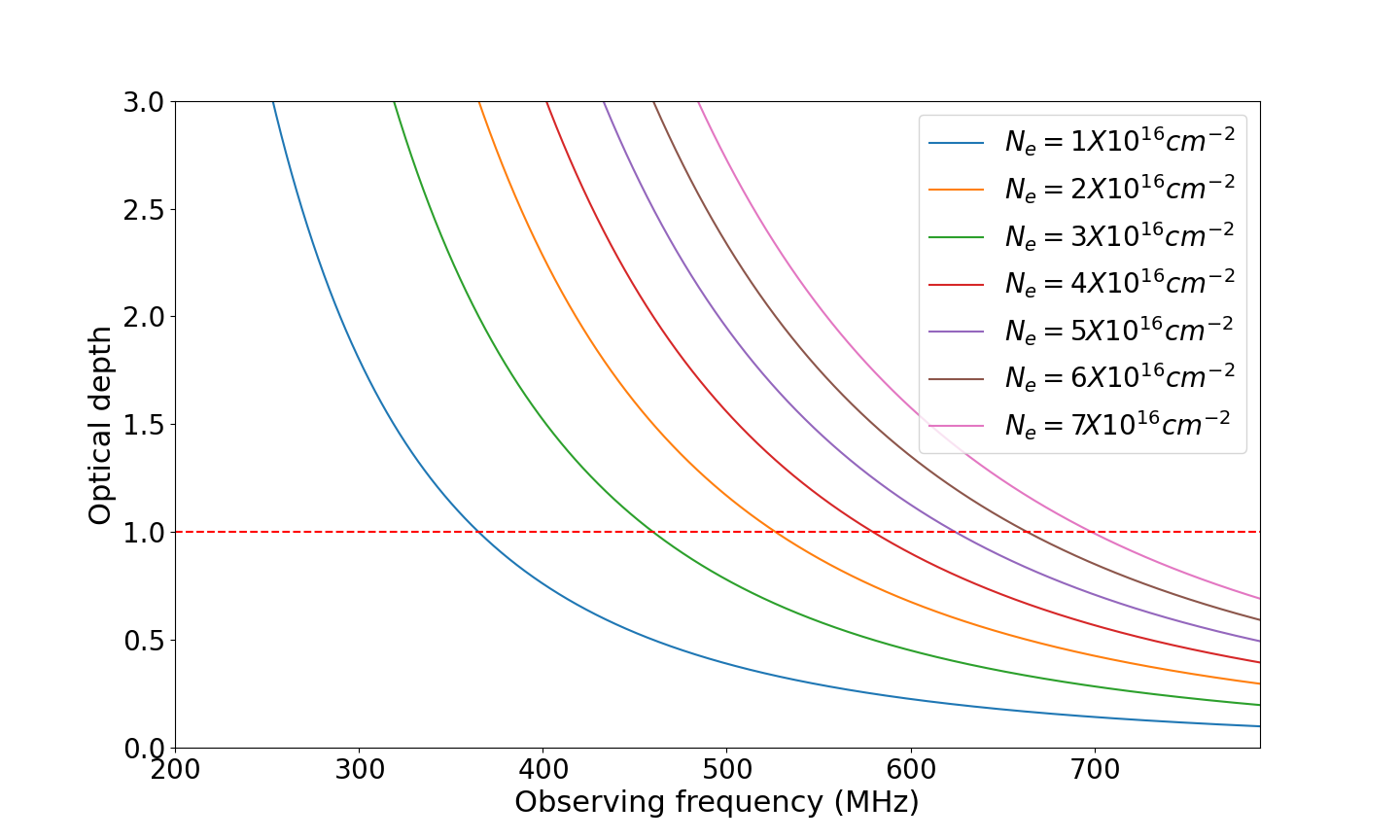}\hfill
 \includegraphics[width=.52\linewidth]{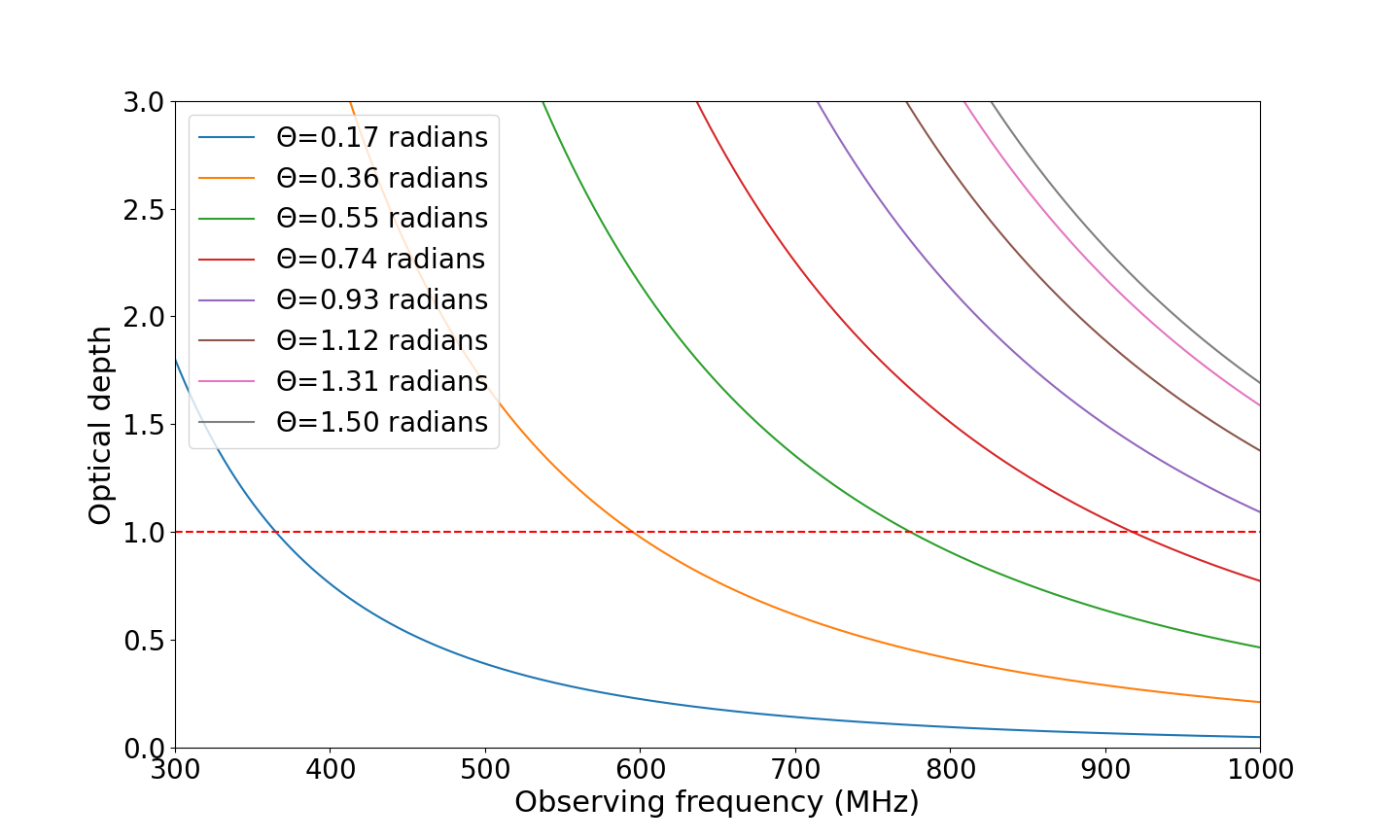}
\caption{The optical depth for synchrotron absorption as the function of frequency with different combinations of the $p$ (upper left panel), $B$ (upper right panel), $N_{e}$ (lower left panel) and $\theta$ (lower right panel) is shown. The red dotted line corresponds to optical depth equals to 1. The intersection of red line with the curves at different values of $p$ (upper left panel, considering $B = 10 \mathrm{Gauss}$, $N_{e} = 10^{16} \mathrm{cm^{-2}}$ and $\theta$ = 0.17 $\mathrm{radians}$), $B$ (upper right panel, considering $p$ = 2, $N_{e} = 10^{16} \mathrm{cm^{-2}}$ and $\theta$ = 0.17 $\mathrm{radians}$), $N_{e}$ (lower left panel, considering $B$ = 10 $\mathrm{Gauss}$, $p$ = 2 and $\theta$ = 0.17 $\mathrm{radians}$) and $\theta$ (lower right panel, considering $B = 10 \mathrm{Gauss}, ~N_{e} = 10^{16} \mathrm{cm^{-2}}$ and $p$ = 2) represents the cut-off frequency. It is evident that even a small change in one of the parameter in the Equation \ref{synchrotron}, can give rise to the change in cut-off frequency.}
\label{cutoff_parameter}
\end{figure*}

If the eclipse medium is magnetised, cyclotron absorption could also be responsible for the observed eclipses. For cyclotron absorption to be a viable eclipse mechanism, temperatures of the order of $10^{7}$ K are required, but the cyclotron approximation is valid for temperatures $\mathrm{T \le 1.9 \times 10^{5} K}$. Therefore, cyclotron absorption is not the major mechanism that explains the observed eclipses. Consequently we concluded that the synchrotron absorption by the free electrons is the major eclipse mechanism for the observed eclipse.
Figure \ref{cutoff_parameter} shows that with even a slight change of the variable parameters in Equation \ref{synchrotron} can manifest in a variation of the observed $\mathrm{\nu_{c}}$. The upper left panel of Figure \ref{cutoff_parameter} indicates the changes in optical depth and hence on the $\mathrm{\nu_{c}}$ with changes in the $\mathrm{p}$, where other parameters in Equation \ref{synchrotron} are kept constant (taking $\mathrm{\theta=0.17 ~ radians,~ B ~= ~10 ~ Gauss ~and~ N_{e} ~= ~10^{16}~ cm^{-2}}$). We note that with the increase in value of $p$ the $\nu_{c}$ is decreasing. We considered $p$ values restricted between 2$-$7 \citep{Dulk&Marsh}. Lower right panel of the Figure \ref{cutoff_parameter} indicates the changes in optical depth and hence on the $\nu_{c}$ with changes in the $\theta$, where other parameters in Equation \ref{synchrotron} are kept constant (taking $ \mathrm{p~ = ~2,~ B ~= ~10 ~Gauss ~and ~N_{e} ~= ~10^{16} ~cm^{-2}}$). With the increase in the value of $\theta$, the $\nu_{c}$ is found to be increasing. Lower left panel of Figure \ref{cutoff_parameter} depicts the changes in optical depth and hence on the $\nu_{c}$ with change in electron density ($n_{o}$) in the eclipse region and keeping other parameters constant in Equation \ref{synchrotron} (taking $\theta ~= ~0.17, ~p ~=~ 2, ~and ~B ~= ~10 ~ \mathrm{Gauss}$). An increase in the value of $N_{e}$ results in increase in the $\nu_{c}$. Similarly, upper right panel of Figure \ref{cutoff_parameter} shows, changes in optical depth and hence on the $\nu_{c}$ with change in magnetic field ($B$), keeping other parameters constant in Equation \ref{synchrotron} (taking $\mathrm{\theta ~= ~0.17, ~ p = 2 ~ and~ N_{e} ~= ~10^{16} ~cm^{-2}}$). An increased value of the magnetic field in the eclipse region, results in an enhanced value for the $\nu_{c}$. 

\section{Calculation of best fit values of $p$ and $\theta$}
\label{Calculation of best fit values of p and theta}
\begin{figure}[!htbp]
\begin{center}
\includegraphics[width=0.4\textwidth,angle=0]{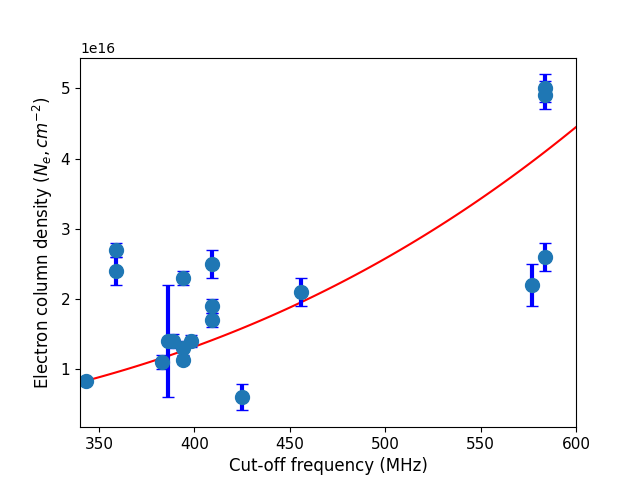}
\caption{The plot of electron column density vs the cut-off frequency considering the eclipses where exact value of both is known. The red curve represents the best fit curve and based on this we get the optimal value of $p$ and $\theta$ in section \ref{Electron column density contribution to eclipse cut-off frequency changes}.}
\label{fig:Ne_cutoff}
\end{center}
\end{figure}

To find the best fit value of $p$ and $\theta$, we created a plot where $N_{e}$ was plotted against $\nu_{c}$ (see Figure \ref{fig:Ne_cutoff}). The theoretical prediction of the  relationship between $N_{e}$ ( where $N_{e}$ = $n_{o} \times L$) and $\nu_{c}$ is described by Equation \ref{equ: Ne_cutoff} (obtained using Equation \ref{synchrotron}). 
\begin{equation}
\label{equ: Ne_cutoff}
    n_{o} = \frac{4m_{e}c\nu^{\frac{p}{2}+2}}{3^{\frac{(p+1)}{2}} \Gamma (\frac{3p+2}{12}) \Gamma(\frac{3p+22}{12}){e^{2}L}\sin \theta^{\frac{p+2}{2}}}
\end{equation}
Assuming $B = ~10 ~\mathrm{Gauss}$, $p$, $\theta$ remain constant in the above equation across different eclipses and taking optical depth ($\tau$) equals to 1, we performed the curve fitting analysis to find the best fit curve and hence the values of $p$ and $\theta$. The best fit curve is shown in Figure \ref{fig:Ne_cutoff} and the corresponding best fit value of $p$ and $\theta$ is given in section \ref{Electron column density contribution to eclipse cut-off frequency changes}. Observing Figure \ref{fig:Ne_cutoff}, it becomes apparent that the fitting is not good, indicating variations in other parameters ($p$, $\theta$, $B$) between different eclipses.

\end{document}

%% file: table1.tex
\begin{table*}[!htb]
\begin{center}
\caption{Summary of the observations.}
\label{tab:Table1}
\vspace{0.3cm}
\label{discovery}
\begin{tabular}{|l|l|l|l|l|l|l|l|}
\hline
Backend      & Frequency (bandwidth) & $T_{res}^{a}$ & $F_{res}^{b}$ &  S$_{min}$ & Span & Epochs &${\sigma_{\tiny  TOA}^{d}}$ \\
&(MHz) &($\mu$s)&(MHz) &(mJy)& & analysed & ($\mu$s)\\
\hline
Legacy GMRT $^{\ast}$  & 306$-$338 (33) & 61.44 & 0.064  & $0.06^{a}$            & 2011$-$2015 & 5 &  $2.6$          \\
\hline
Legacy GMRT $^{\ast}$ & 591$-$623 (33) & 61.44 & 0.064  & $0.05^{a}$            & 2014$-$2017 & 2 &  $6.0$          \\
\hline

Upgraded GMRT (band-3) $^{\ast\ast}$ &300$-$500 (200)& 81.92 & 0.048  & $0.02^{c}$  &  2018$-$2022 & 31 & $1.0$        \\
\hline
Upgraded GMRT (band-4)$^{\ast\ast}$ &550$-$650 (200) & 81.92 & 0.048 & $0.02^{c}$  &  2018$-$2022 & 22 & $2.5$        \\
\hline
\end{tabular}
\end{center}
$^{\ast}$ : \cite{roy2010real}\\
$^{\ast\ast}$ : \cite{gupta2017upgraded} \\
$^{a}$: Time resolution\\
$^{b}$: Frequency resolution\\
$^{a}$: 5$\sigma$ detection sensitivity calculated using the radiometer equation \citep{LorimerKramer} considering  gain of 0.32 K/Jy for the legacy GMRT system, system temperature $\sim$ 108 K at 322 MHz and $\sim$ 92 K at 607 MHz, 20\% duty cycle, 26 antennas ($\sim $ number of antennas used in the observations), considering 2 polarisation's and 60 mins of observing time in coherent array mode.\\
$^{c}$: 5$\sigma$ detection sensitivity calculated using the radiometer equation \citep{LorimerKramer} considering  gain of 0.38 K/Jy for the upgraded GMRT system at band-3 and 0.35 K/Jy for the upgraded GMRT at band-4, system temperature $\sim$ 123 K at band-3 including the sky temperature in the direction of J1544+4937 and $\sim$ 106 K at band-4, 20\% duty cycle, 26 antennas ($\sim $ number of antennas used in the observations), considering 2 polarisation's and 60 mins of observing time in coherent array mode.\\
$^{d}$: Error in TOA estimation calculated using the radiometer equation considering the above parameters for GMRT in 10 mins of observing time.\\

\vspace{1cm}
\end{table*}

%% file: table3.tex
\begin{table*}[!htb]
\begin{center}
\caption{Table listing the cut-off frequency values for multiple observing epochs spanning a decade for PSR J1544+4937 along with corresponding \textbf{non-eclipse phase flux density} and electron column density.}
\label{tab:Table3}
\vspace{0.3cm}
\label{table2}
\begin{tabular}{|l|l|l|l|l|l|l|l|}
\hline
Epoch   & Cut-off  ($\nu_{c}$)$^{\dagger}$ & Orbital $^{\ast\ast}$ & Observed $N_{e}$ $^{b}$ & Flux   & Flux & $2R_{E}$ & $\Dot{M_{c}}$\\
& frequency  &phase & $(cm^{-2})$ & density $^{\dagger\dagger}$  & density & (\(R_\odot\))& ($10^{-14}$ \(M_\odot\) $yr^{-1}$)\\
& (MHz)&  &  & (band-3, mJy)  & (band-4, mJy) & & \\
\hline
05 June 2011   & $>$ 338 & 0.30   & ($8.1 \pm 14) \times 10^{14\ast} $  & 0.729 & - & 1.27 & 2.24\\
\hline
17 November 2012   & $>$ 338    &  0.19   & $(5.6 \pm 0.5)\times 10^{15\ast}$ & 3.819 & - & 0.58 & 1.85\\
\hline
09 July 2014   & $>$ 338   & 0.18       & $(7.0 \pm 11.9) \times 10^{14\ast}$ & 1.98 & - & - & -\\
\hline
01 November 2014 & $>$ 624    & 0.28    & (2.9 $ \pm 0.2) \times 10^{16\ast}$  & - & 0.75 & -& -\\
\hline
21 April 2015    & $>$ 338   & 0.18  & (1.1 $\pm 0.08) \times 10^{16\ast}$  & 1.101  & - & - & - \\
\hline 
02 June 2015     &  $>$ 338   & 0.30        &  (8.0 $\pm 0.7) \times 10^{15\ast}$ & 1.75 & - & - & - \\
\hline
07 March 2017    & $>$ 624    & 0.26        &  (1.0 $\pm 0.1) \times 10^{16\ast}$  & - & 0.79 & - & - \\
\hline
06 February 2018  & \bf{389 ± 7}    & 0.25         & ($1.4 \pm 0.1) \times 10^{16}$  & 2.43 & - &0.65 & 3.44\\
\hline
17 April 2018    &  \bf{409 ± 7}     & 0.25        & $(2.5 \pm 0.2) \times 10^{16}$  & 1.229 & 0.42 & 1.22 & 10.56\\ 
\hline
07 May 2018      &  \bf{359 ± 7}     & 0.25         & $(2.4 \pm 0.2) \times 10^{16}$    & 1.569 & - & 0.46 & 2.24\\
\hline
02 December 2019  & $\ge$ 480     & 0.26          & $(1.1 \pm 0.2) \times 10^{16\ast}$  & 0.342 & - & 1.01 & 5.93 \\
\hline
10 January 2020   & \bf{398 ± 7}      &  0.25         & $(1.4 \pm 0.08) \times 10^{16}$  & 0.922 & - & 0.58 & 2.88\\
\hline
15 February 2020    & \bf{383 ± 7}     & 0.23          & $(1.1 \pm 0.1) \times 10^{16}$ & 0.936 & - & - & - \\
\hline
16 June 2020   & \bf{359 ± 7}     &  0.24        & ($2.7 \pm 0.1) \times 10^{16}$    & 1.147 & - & 0.59 & 3.48\\
\hline
26 June 2020   &  $<$ 550    & 0.24          & $(3.0 \pm 0.2) \times 10^{16}$   & - & 0.53 & - &-\\
\hline
29 June 2020  & $>$ 480       & 0.28         & $(2.5 \pm 0.6) \times 10^{15\ast}$  & 0.858 & 0.42 & 1.46 & 4.9\\
\hline
13 July 2020   & \bf{343 ± 7}      & 0.25         & $(8.3 \pm 0.2) \times 10^{15}$  & 0.794 & - & - & -\\
\hline
07 September 2020    & $>$ 480      &  0.30   & (7.9 $\pm 0.6) \times 10^{15\ast}$ &  0.427 & - & 2.98 & 25.3\\
\hline
08 October 2020  & \bf{386 ± 7}     & 0.26          & $(1.4 \pm 0.8) \times 10^{16 }$ & 1.165 & - & 0.79 & 4.44\\
\hline
12 February 2022  & $>$ 740  &  0.28        & $(4.3 \pm 0.05) \times 10^{16\ast}$  & 0.862 & 0.53 & 1.04 & 12.15\\
\hline
28 June 2022  &  $>$ 740   & 0.31          & (1.0 $\pm 0.06) \times 10^{16\ast}$  &  0.593 & 0.64 & - & -\\
\hline
08 July 2022 &  $>$ 470, $<$ 560 &  0.26   & ($2.8 \pm 0.3) \times 10^{16 }$   & 1.392 & 0.73  & 1.02 & 8.72\\
\hline
23 September 2022 &  $>$ 740     & 0.30  & (7.0 $\pm 1.2) \times 10^{15\ast}$ & 0.915  & 0.29 & 1.85 & 11.67\\
\hline
30 September 2022   &  $>$ 740    & 0.30 & (6.1 $\pm 1.7) \times 10^{15\ast}$  & 0.452  & 0.14 & - & - \\
\hline
01 October 2022    & \bf{577 ± 7} & 0.23   & $(2.2 \pm 0.3) \times 10^{16}$ & 1.102 & 0.58 & 1.68 & 16.8\\
\hline
13 November 2022    &  $>$ 480, $<$ 560  & 0.27    & ($1.7 \pm 0.1) \times 10^{16}$ & 0.754 & 0.74 & - & -\\
\hline
25 November 2022    &  \bf{584 ± 7}  & 0.25   & $(2.6 \pm 0.2) \times 10^{16}$  &  1.327 & 0.82 &1.13 & 9.72\\
\hline
13 December 2022 & \bf{394 ± 7}  & 0.29          & ($1.1 \pm 0.06) \times 10^{16}$ & 2.36 & - & 0.78 & 3.83\\
\hline
24 December 2022\_1  $^{a}$ & \bf{584 ± 7}  & 0.27  & ($4.9 \pm 0.2) \times 10^{16}$ & 2.281 & 1.03 &0.84  & 9.56\\
\hline
24 December 2022\_2  $^{a}$ & \bf{584 ± 7}    & 0.27  & ($5.0 \pm 0.2) \times 10^{16}$ & 1.285 & 0.94 & 0.99 & 12.67\\
\hline
30 December 2022\_1 $^{a}$ & \bf{456 ± 7} &  0.27  & ($2.1 \pm 0.1) \times 10^{16}$ & 1.405 & 1.12 & - & - \\
\hline
30 December 2022\_2 $^{a}$& \bf{409 ± 7}  & 0.27  & ($1.7 \pm 0.2) \times 10^{16}$ & 0.907 & 0.443 & 1.12 & 7.93\\
\hline
11 January 2023  & \bf{394 ± 7}   & 0.28          & $(2.3 \pm 0.1) \times 10^{16}$ & 1.653 & 1.31 & 0.90 & 7.15\\
\hline
14 January 2023 & \bf{394 ± 7}  & 0.27          & ($1.3 \pm 0.1) \times 10^{16}$ &  1.924 & 0.86 & 0.67 & 3.06\\  
\hline
28 January 2023\_1 $^{a}$ & \bf{409 ± 7}  &   0.28    & ($1.9 \pm 0.1) \times 10^{16}$ & 1.76 & 1.12 & 0.72 & 4.14\\ 
\hline
28 January 2023\_2 $^{a}$ & -  &   0.28    & ($1.9 \pm 0.2) \times 10^{16}$ & 2.098 & 0.71 & 0.90 & 6.5\\ 
\hline
10 February 2023 & $>$ 470, $<$ 560 &   0.27       & $(2.8 \pm 0.3) \times 10^{16}$ & 1.421 & 0.76 & 0.81 & 5.46\\ 
\hline
04 March $2023^{c}$ &  \bf{425 ± 7} &      -    &  - & 1.425  & 0.53 &1.06 & -\\
\hline
26 March 2023 &  740 &   0.28    & $(2.2 \pm 0.4) \times 10^{16\ast}$ & 0.932 & 0.77 &1.13 & 9.83\\

\hline
\end{tabular}
\end{center}

$^{\dagger}$ : The observed $\nu_c$. The values in bold depict the epochs where the exact value of $\nu_c$ is known. \\
$^{\ast\ast}$ : The orbital phase at which the maximum value of $N_{e}$ is obtained \\
$^{b}$: The observed $N_{e}$ at the corresponding orbital phase. The values in bold depict the epochs where maximum $N_e$ is exactly known in the eclipse medium.\\
$^{\dagger\dagger}$: Non-eclipse phase flux densities at band 3 computed using the standard gain of the GMRT antennas \\ 
$^{\ast}$: The corresponding values of the observed $N_{e}$ are the lower limits \\
$^{a}$ : Observations spanning two consecutive binary orbits for PSR J1544+4937 \\
$^{c}$ : $N_{e}$ is not calculated as the TOAs for this eclipse were irregular in the eclipse phase. 

\vspace{1cm}
\end{table*}